\documentclass[%
 aip,
 amsmath,amssymb,
 reprint,%
]{revtex4-1}

\usepackage{graphicx}
\usepackage{dcolumn}
\usepackage{bm}

\usepackage[utf8]{inputenc}
\usepackage[T1]{fontenc}
\usepackage{mathptmx}
\usepackage{etoolbox}

\usepackage{color}
\usepackage{cleveref}
\crefformat{figure}{#2Fig.~#1#3}
\crefformat{equation}{#2Eq.~(#1)#3}
\crefformat{table}{#2Table~#1#3}
\crefformat{section}{#2Section~#1#3}

\makeatletter
\def\@email#1#2{%
 \endgroup
 \patchcmd{\titleblock@produce}
  {\frontmatter@RRAPformat}
  {\frontmatter@RRAPformat{\produce@RRAP{*#1\href{mailto:#2}{#2}}}\frontmatter@RRAPformat}
  {}{}
}%
\makeatother
\begin{document}

\preprint{AIP/123-QED}

\title[Thermal conductivity suppression in uranium-doped thorium dioxide due to phonon resonant scattering]{Thermal conductivity suppression in uranium-doped thorium dioxide due to phonon resonant scattering}
\author{Zilong Hua}
    \email{zilong.hua@inl.gov}
    \affiliation{Materials Science and Engineering Department, Idaho National Laboratory, Idaho Falls, ID 83415, USA}
\author{Saqeeb Adnan}
    \affiliation{Department of Mechanical and Aerospace Engineering, The Ohio State University, Columbus, OH 43210, USA}
\author{Amey R. Khanolkar}
    \affiliation{Materials Science and Engineering Department, Idaho National Laboratory, Idaho Falls, ID 83415, USA}
\author{Karl Rickert}
    \affiliation{KBR, Dayton, OH 45431, USA}
\author{David B. Turner}
    \affiliation{Azimuth Corporation, Fairborn, OH 45324, USA}
\author{Timothy A. Prusnick}
	\affiliation{KBR, Dayton, OH 45431, USA}
\author{J. Matthew Mann}
	\affiliation{Air Force Research Laboratory, Sensors Directorate, Wright-Patterson AFB, OH 45433, USA}
\author{David H. Hurley}
    \affiliation{Materials Science and Engineering Department, Idaho National Laboratory, Idaho Falls, ID 83415, USA}
\author{Marat Khafizov}
    \email{khafizov.1@osu.edu}
	\affiliation{Department of Mechanical and Aerospace Engineering, The Ohio State University, Columbus, OH 43210, USA}
\author{Cody A. Dennett}
	\affiliation{Materials Science and Engineering Department, Idaho National Laboratory, Idaho Falls, ID 83415, USA}
	\affiliation{Department of Nuclear Science and Engineering, Massachusetts Institute of Technology, Cambridge, MA 02139, USA}

\date{\today}

\begin{abstract}
In this work, the thermal transport properties of thorium dioxide (ThO$_2$, thoria) with low levels of substitutional uranium (U) doping are explored. We observe strong indications of resonant phonon scattering, an interaction between phonons and electronic degrees of freedom, induced by this doping in addition to common ``impurity'' scattering due to mass and interatomic force constant differences. Uranium doping levels of 6\%, 9\%, and 16\% were studied in a single hydrothermally synthesized U-doped thoria crystal with spatially-varying U doping levels. Within this crystal, isoconcentration regions with relatively uniform doping were located for local thermal conductivity measurements using a thermoreflectance technique. The measured thermal conductivity profiles in the temperature range of 77--300~K are compared to predictions of an analytical Klemens-Callaway thermal conductivity model to identify impacts from different phonon scattering mechanisms. Highly suppressed thermal conductivity at cryogenic temperatures at these doping levels suggests that phonon resonant scattering plays an important role in thermal conductivity reduction in U-doped thoria.
\end{abstract}

\maketitle

\section{\label{sec:level1}Introduction}

Dopants play a key role in thermal transport in fluorite oxides across a range of energy applications including solid oxide fuel cells~\cite{Mogensen2000}, thermal barrier coatings~\cite{pan2012low}, laser host materials, thermoelectrics~\cite{zhao2016ultrahigh}, and nuclear fuels~\cite{hurley2022thermal}. In oxide materials, thermal energy is primarily conducted by lattice vibrations, namely, phonons~\cite{Berman1976heatconduction}. Dopants acting as point defects (substitutional impurities or interstitials) scatter phonons, suppressing thermal conductivity~\cite{walker1963phonon}. Traditionally, such point-defect-induced thermal conductivity suppression in the actinide fluorite oxides has been described using a Rayleigh scattering formulation~\cite{gibby1971effect, ohmichi1981relation, klemens1955scattering, abeles1963lattice}. However, for certain point defects, Rayleigh’s formulation fails to explain low-temperature thermal transport behavior due to phonon interaction with localized vibrational modes~\cite{Slack1961, Pohl1962, Slack1964,kundu2019effect}. In the case of paramagnetic impurities, the interaction between the point defect impurity and the surrounding host lattice, through modulation of the crystalline electric field, induces a spin transition by perturbing the orbital motion of paramagnetic electrons. The spin transition energy is imparted to the host lattice leading to a localized (resonant) mode that gives rise to phonon resonant scattering or phonon-spin scattering~\cite{maradudin1966pt2,Mattuck1960}. Phonon resonant scattering has a characteristic effect on low temperature thermal conductivity, tending to introduce a dip in thermal conductivity where one typically expects maximum thermal conductivity with Rayleigh-type scattering only. In some cases, the thermal conductivity reduction induced by phonon resonant scattering can be orders of magnitude higher than that induced by Rayleigh-type scattering~\cite{Begheri2020}. 

Thoria (ThO$_2$) has been considered a promising candidate material for advanced nuclear fuel cycles due to its high melting point, reasonable thermal conductivity, and relatively low radioactive waste footprint~\cite{Herring2001}. Recently, a number of works have focused on better understanding the thermal and mechanical properties of thoria and the possible changes of these properties induced by defects generated in extreme nuclear reactor environments. The use of high-quality single crystal thoria has been the critical foundation of these works, as desired microstructure features can thus be introduced in isolation~\cite{hurley2022thermal}. Specifically, thermal conductivity, and corresponding changes induced by point defects and dislocation loops, have been investigated by Dennett, Deskins, Jin, and coworkers~\cite{Dennett2020ThO2, deskins2021thermal, Dennett2021ThO2, Jin2022, Deskins2022Combined}, by combining the efforts of advanced experimental tools (laser-based thermoreflectance and transmission electron microscopy) and state-of-the-art modeling approaches (linearized Boltzmann transport equation solutions with inputs from first principles calculations, defect evolution models, and advanced electron microscopy characterization). 
    
In this work, we further explore how resonant phonon scattering impacts the thermal conductivity of thoria by doping single crystal thoria with uranium (U). Here, local thermal conductivity ($k$) in a temperature range ($T$) of 77--300~K was measured on a U-doped thoria sample. By using a hydrothermal synthesis technique, varied U doping was introduced in isolated spatial regions from a single crystal thoria seed. Three isoconcentration regions with uniform U-doping percentages (6\%, 9\%, and 16\%) were identified using Raman spectroscopy and X-ray fluorescence (XRF), and then fiducially marked using a focused-ion-beam (FIB) for thermoreflectance measurements. The extracted $k$--$T$ curve is compared with that calculated from a Klemens-Callaway thermal transport model to isolate the impacts of Rayleigh-type phonon-point defect scattering and phonon resonant scattering on thermal conductivity. We find that phonon resonant scattering contributes more than Rayleigh-type scattering to the thermal conductivity reduction in this temperature range, and the correlated intensity is not linear with the doping level. This work shows the profound effect that introducing atoms with 5$f$ electrons has on the thermal transport behavior of fluorite oxides.

\section {Materials and Methods}

\subsection{Crystal Preparation}

Crystal synthesis followed the previously published procedure for hydrothermal synthesis of (U,Th)O$_2$, with a feedstock composed of 0.2~g of UO$_2$ (IBI Labs, 99.99) and 2.1~g of ThO$_2$ (IBI Labs 99.99)~\cite{Rickert2022}. The bottom 127~mm of the silver growth ampoule was heated to 650°C over 4.5~hrs. The top 139.7~mm was heated to 525$^\circ$C over 4.5~hrs, paused at 525$^\circ$C for 18~hrs, and then heated to 600$^\circ$C over 2~hrs. These temperatures were held for 45 days at a pressure of 23.5~kpsi before being cooled to 10$^\circ$C over 24~hrs. The product was bisected with a water-cooled diamond wire saw (STX-202P, MTI Corporation) to obtain halves with the largest surface area. The cut face was polished on a rotary disc polisher with diamond laps, ending with a 0.1~{\textmu}m grit size.  

\subsection{Isoconcentration Region Analysis}

\begin{figure*}
    \centering
    \includegraphics[width=0.7\textwidth]{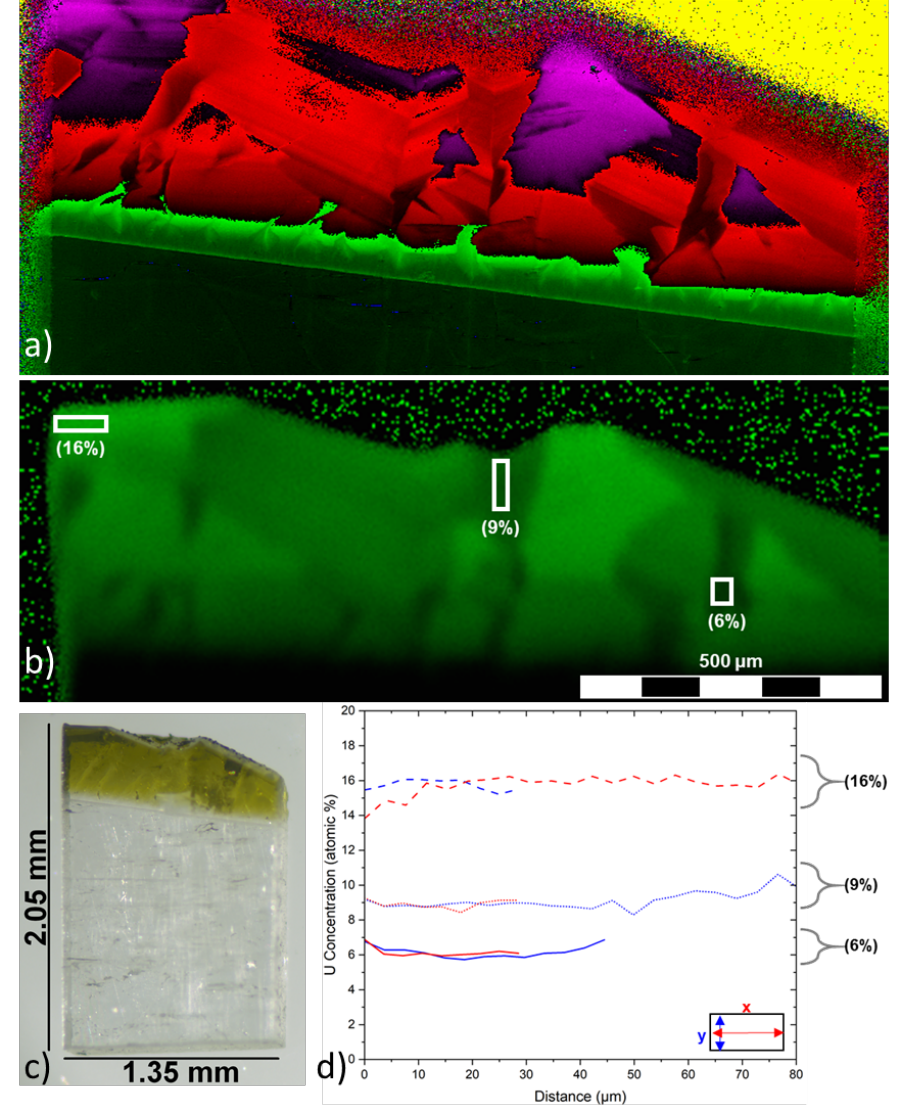}
    \caption{(a) {\textmu}-Raman map showing spatial variation of $T_{2g}$ FWHM of the U$_x$Th$_{1-x}$O$_2$ region and ThO$_2$ seed (purple = broad peak indicating more defects/doping, dark green = narrow pure thoria peak); (b) XRF map of the same showing U signal intensity; (c) Optical image of the entire as-prepared crystal; and (d) XRF line scans showing atomic composition across the isoconcentration regions drawn in (b).\label{fig:sample}}
\end{figure*}

{\textmu}-Raman spectroscopy and XRF were used to identify different isoconcentration regions for thermal transport analysis and to measure the as-grown uranium concentrations, respectively. The Raman and XRF maps, an optical image of the as-polished crystal, and XRF line scans of the isoconcentration region are shown in \cref{fig:sample}. A summary of the Raman and XRF analyses are provided here and further details can be found in the Supplementary Materials.

{\textmu}-Raman measurements were carried out using a Renishaw Invia Reflex Raman microscope with a 633~nm excitation source equipped with 1200~l/mm dispersion grating and a standard Renishaw Si CCD~\cite{Rickert2019}. A Si calibration was performed prior to the measurements. The laser was focused through a 50$\times$ long-working-distance objective with a numerical aperture (NA) of 0.5. A {\textmu}-Raman map of the polished side of the growth region was performed with a single 0.5~sec exposure for each point and points were collected in 2~{\textmu}m by 2~{\textmu}m intervals. The laser focus was manually adjusted every 50~{\textmu}m and the intervening focal points were extrapolated. Renishaw Raman spectroscopy software WiRE was used to calculate and plot the signal to baseline intensity and full peak width at half of the maximum intensity (FWHM) for the $T_{2g}$ peak.

XRF data were collected using two large-area silicon drift detectors set to a 40~keV range on a Bruker M4 TORNADO$^{\text{PLUS}}$, with an Rh source set at 50~kV and 300~{\textmu}A, and a spot size of 14.7~{\textmu}m (independently measured on this specific instrument). An XRF map of the polished side of the growth region was taken and spots were analyzed every 4~{\textmu}m (in both x and y directions). The dwell time was 50~ms/pixel, with each pixel being 4~{\textmu}m in diameter, and 15 cycles were performed and averaged together. A Zr calibration was performed prior to measurements being taken. Visual inspection of the overall XRF spectrum of each map was used to initially determine which elements may be present. In all cases, peaks that did not match U or Th did not reasonably match any other elements, thus only U and Th were included in the quantification step. 

\subsection{Thermal Transport Measurements}

Local thermal transport was measured using a spatial-domain thermoreflectance (SDTR) technique \cite{Hua2012,Hurley2015}. In SDTR, a continuous-wave (CW) laser (Coherent OBIS 660~nm) with a periodically modulated intensity is used to locally heat the sample and induce a thermal wave. The propagation of the thermal wave is detected using a CW laser with constant intensity (Coherent Verdi 532~nm) through the thermoreflectance effect~\cite{Hua2012}. Thermal transport properties, such as thermal conductivity and thermal diffusivity, can be extracted using a thermal diffusion model and corresponding boundary conditions~\cite{Hua2012, maznev1995thermal, Hurley2015}. In order to improve the measurement spatial resolution, laser beams are focused using a 50$\times$ long-working-distance objective lens (Olympus SLMPlan 50$\times$). The spot size on the sample surface is $\sim$1~{\textmu}m for each laser, with the power of $\sim$2~mW and $\sim$0.3~mW for the heating and probe lasers, respectively.
    
Thermal transport property measurements were conducted in the isoconcentration regions and in the seed ThO$_2$ area over a 77--300~K temperature range with a step of 25~K and temperature fluctuation of less than 3~K~\cite{Dennett2021ThO2}.  At each temperature, at least 6 sets of measurements at four modulation frequencies, 10, 20, 50, and 100~kHz, were performed to statistically reduce the uncertainty of measured thermal conductivity. A 63~nm film of gold was sputter-coated on the sample surface to improve the thermoreflectance effect and energy absorption at this heating laser wavelength~\cite{Hatori2005}. This film thickness was chosen based on a sensitivity analysis to ensure independent extraction of thermal conductivity ($k$) and thermal diffusivity ($D$) at room temperature. At room temperature, the SDTR thermal model can be validated by calculating the heat capacity ($C_p$) from measured $k$ and $D$ and comparing it to reference measurements made on crystals using the same growth process~\cite{Dennett2020ThO2}. At low temperature, $D$ was the sole parameter optimized from SDTR measurements to improve accuracy; $k$ was then calculated at each temperature as $k=D\rho C_p$, where $C_p$ and density ($\rho$) for pure ThO$_2$ was obtained from previously reported values \cite{Dennett2020ThO2}. In U-doped regions, $C_p$ of pure ThO$_2$ was used as the estimation of $C_p$ values for U-doped ThO$_2$ as the heat capacity of pure fluorite UO$_2$ varies little from that of ThO$_2$ in this temperature range~\cite{hurley2022thermal}.

\begin{figure*}
    \centering
    \includegraphics[width=1\textwidth]{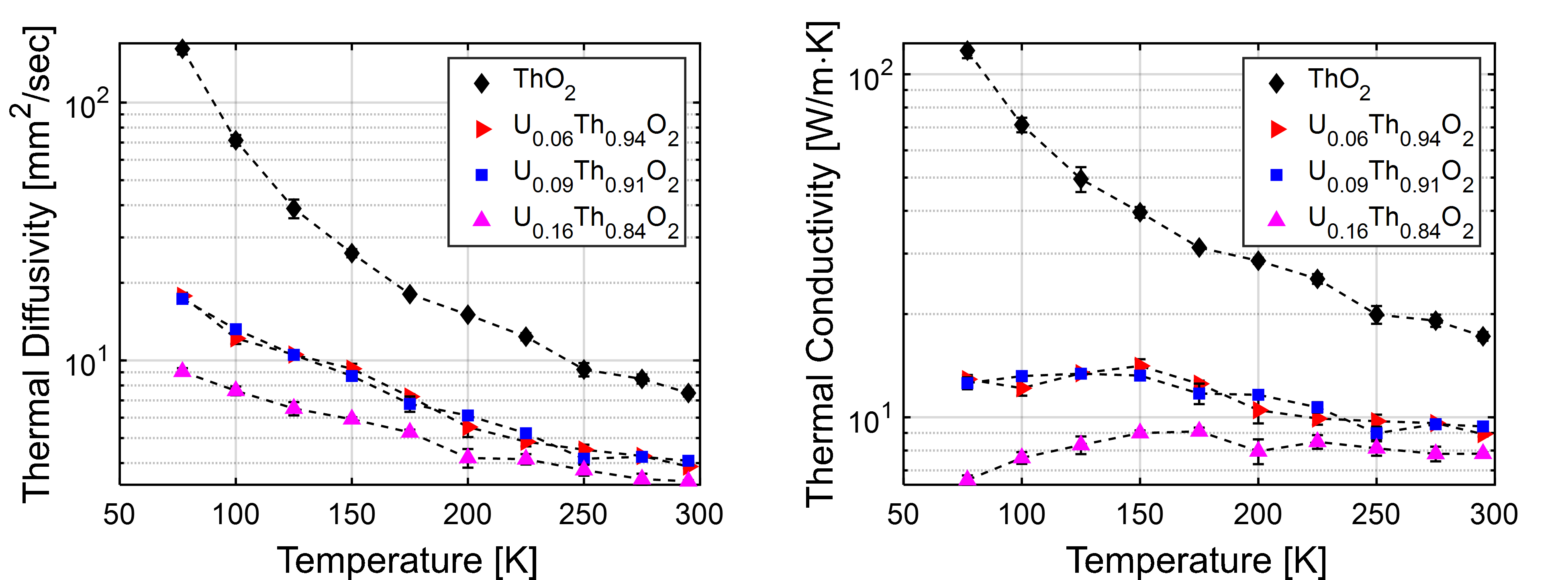}
    \caption{Thermal diffusivity (left) and conductivity (right) measured using SDTR in the ThO$_2$ seed crystal, and isoconcentration regions of 6\%, 9\%, and 16\% uranium doping in ThO$_2$.\label{fig:expdata}}
\end{figure*}

\section{Results and Discussion}

The measured thermal diffusivity ($D$) and calculated thermal conductivity ($k$) in the 6\%, 9\%, and 16\% U doped ThO$_2$ regions, and in the seed ThO$_2$ region, of the heterogeneous crystal are plotted in \cref{fig:expdata}. The standard errors determined on the basis of multiple measurements at each concentration and temperature are $\sim$5\% for the majority of concentrations and temperatures. In the seed ThO$_2$ at 125~K, the uncertainty is slightly higher ($\sim$10\%) due to the high diffusivity at low temperatures and challenges associated with stabilizing this temperature using LN$_2$ coolant. In the seed ThO$_2$, $k$ at 77~K is measured as 117~W/m$\cdot$K, comparable to the highest reported value of $k$ for single crystal ThO$_2$ at the same temperature~\cite{Mann2010}. This suggests low impurity levels in the seed crystal. After doping, there are significant reductions in both $k$ and $D$ over the entire temperature range. The relative change in $k$ increases as the temperature decreases. A slight hump is observable in the $k$--$T$ curve between 125--175~K in all U-doped regions. This feature is not reflected in the ThO$_2$ seed, nor is it expected based on previous measurements of hydrothermally-grown ThO$_2$ single crystals~\cite{Mann2010,Dennett2021ThO2,Xiao2022}. The thermal conductivity of 9\% U doped ThO$_2$ does not differ noticeably from that at 6\% U-doping: 12.6~W/m$\cdot$K for the 9\% doped region versus 12.9 W/m$\cdot$K for the 6\% doped region at 77~K. However, with additional doping to 16\%, a significant thermal conductivity reduction was observed: 6.6 W/m$\cdot$K at 77~K, or $\sim$50\% lower than the 9\% doped region. 

\begin{figure*}
    \centering
    \includegraphics[width=1\textwidth]{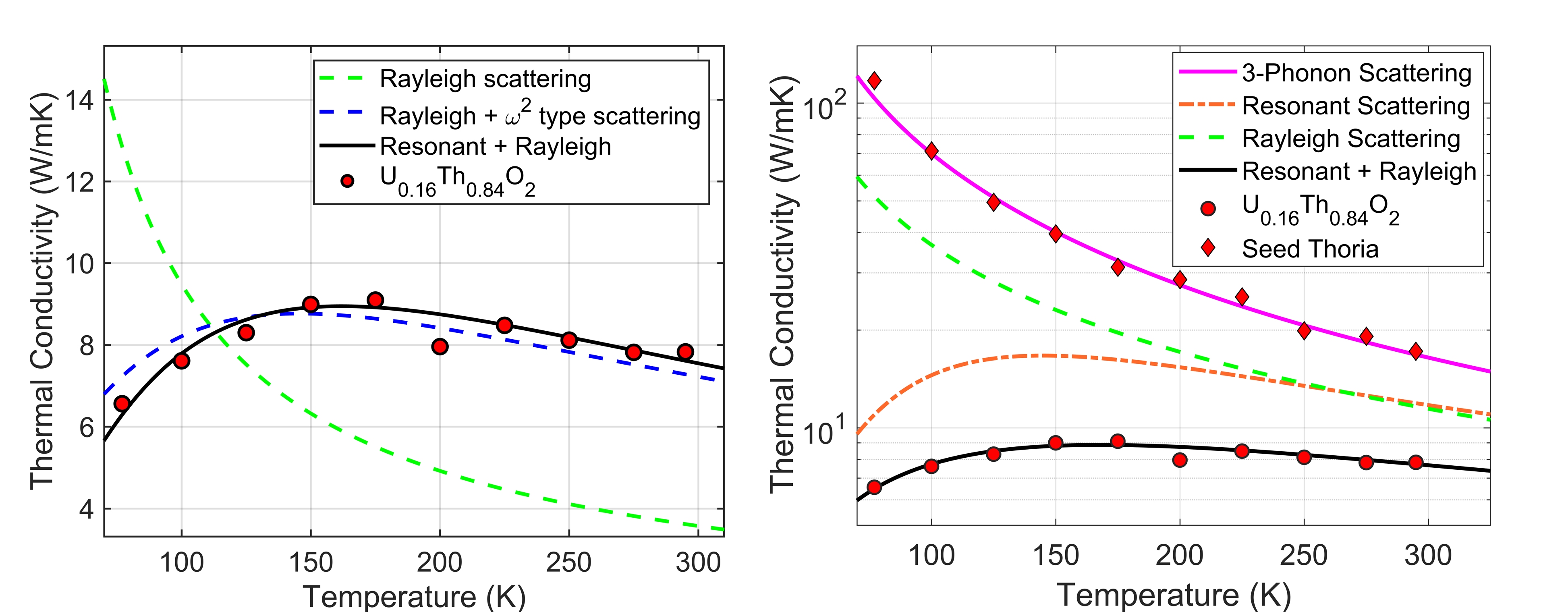}
    \caption{Measured thermal conductivity of 16\% U-doped thoria sample at low temperatures. (Left) Best fit curves using the KCM model with select scattering mechanisms: green dashed line -- fit using only Rayleigh type point defect scattering; dashed blue line -- fit including contribution from $\omega^2$ type scattering mechanism along with Rayleigh scattering; solid black line -- fit considering resonant scattering in combination with Rayleigh scattering. 
    (Right) Best-fit KCM model for 16\% U-doped thoria showing the contributions of Rayleigh and resonant scatterings in comparison to the seed thoria crystal.}
        \label{fig:fitting_options}
\end{figure*}

We analyze our results using a Klemens-Callaway model (KCM) to investigate the impact of individual phonon scattering mechanisms on thermal conductivity. Using Debye’s linear approximation for phonon dispersion, thermal conductivity of our U doped ThO$_2$ system is calculated as
\begin{equation}
    \label{eq:conductivity}
	k= \frac{1}{2\pi^2v^3}\int_0^{\omega_D} \frac{C(\omega,T)v^2\omega^2}{\tau^{-1}(\omega,T)} \,d\omega
\end{equation}
where $\omega$ is phonon frequency, $T$ is temperature, $\omega_D$ is the Debye frequency, and $v$ is the sound velocity~\cite{callaway1959model,chauhan2021indirect}. The Debye sound velocity is expressed through the longitudinal ($v_L$) and transverse ($v_T$) components as $v=(1/3v_L^3+2/3v_T^3)^{- 1/3}$. The sound velocity components $v_L$ and $v_T$ are obtained through elastic stiffness tensor elements $C_{11}$ and $C_{44}$, as $v_L=\sqrt{C_{11}/\rho}$ and $v_T=\sqrt{C_{44}/\rho}$. Values of the elastic stiffness tensor components for pure ThO$_2$ are used as reported in \cite{mathis2022generalized} and the Debye sound velocity was calculated to be 3165~m/s. The Debye frequency is calculated as $\omega_D=v(6N\pi^2/V_0)^{1/3}$, where $N=3$ is the number of atoms in the fluorite unit cell, and $V_0=a^3/4$ is the volume of the unit cell, where $a=5.529$~\AA ~is the lattice constant of pure ThO$_2$~\cite{khafizov2014thermal}. The specific heat, $C(\omega,T)$ is expressed as:
\begin{equation}
    \label{eq:heatcapacity}
	C(\omega,T)= \frac{k_Bx^2e^x}{(1-e^x)^2}
\end{equation}
where $x=\hbar\omega/k_BT$, $k_B$ is the Boltzmann constant, and $\hbar$ is the reduced Planck’s constant. The scattering rate $\tau^{-1}$ is a combination of multiple scattering processes that are summed using Matthiessen's rule as
\begin{equation}
    \label{eq:scatteringrate}
    \begin{split}
	\tau^{-1} = & BT\exp\left(\frac{-T_D}{3T}\right)\omega^2 + A_i \omega^4\\ & + \frac{V_o}{4\pi Nv^3}\Gamma\omega^4+\frac{C_r\omega^4}{(\omega^2-\omega_0^2)^2}F(\omega_0,T)
	\end{split}
\end{equation}
where each term in Eq.~\ref{eq:scatteringrate} represents an individual scattering process \cite{walker1963phonon,chauhan2021indirect,ziman2001electrons}. The first term corresponds to three-phonon scattering, where $T_D=\hbar\omega_D/k_B$ is the Debye temperature. The second term quantifies the contribution from impurities in the pristine sample, described here by Rayleigh-type scattering of phonons as has been successful in the past for ThO$_2$~\cite{Dennett2021ThO2,Deskins2022Combined}. These two scattering processes exist in both seed thoria and U-doped thoria, and the corresponding linear parameters can be obtained by fitting the measured $k$--$T$ profile of seed thoria to the KCM using only the first two terms. These optimized parameters, $A_i=6.5\times10^{-44}$~s$^3$ and $B=3.1\times10^{-18}$~s/K, are then held constant in further model fitting of U-doped thoria.

The scattering mechanisms attributed to U-doping are represented by the last two scattering terms. The third term in Eq.~\ref{eq:scatteringrate} describes phonon scattering by typical point defects using a general Rayleigh scattering expression with $\omega^4$ dependence~\cite{khafizov2019impact, klemens1955scattering}. Here, $\Gamma$ is the point defect scattering parameter, proportional to defect concentration~\cite{klemens1955scattering, deskins2021thermal}. The dashed green line in~\cref{fig:fitting_options} (left) represents the best-fit KCM for 16\% U-doped thoria with only the Rayleigh-type defect scattering term included in \cref{eq:scatteringrate}. This functional form is qualitatively unable to capture the trend apparent in the experimental results, shown as red circles. The dashed blue line considers an alternate point defect scattering functional with an $\omega^2$ dependence as an empirical attempt to fit the data. As can be seen, such an empirical attempt is able to better capture the experimental trend, though some qualitative deviation is still present, particularly at low temperatures. 

A number of alternative approaches to capture additional scattering mechanisms when insulators are doped have been applied with success in the past~\cite{maradudin1966theoretical,walker1963phonon}. We found best agreement with experiment when the additional scattering is represented by the last term in Eq.~\ref{eq:scatteringrate}. This expression has been commonly used to capture phonon resonance scattering and accounts for the time-dependent oscillating harmonic perturbation due to impurities or electronic degrees of freedom 
that resonate with phonons. Here $\omega_0$ is the resonant frequency and $F(\omega_0,T)=(1-e^{-\hbar\omega_0/k_B T})/(1+e^{-\hbar\omega_0/k_BT})$ is a temperature-dependent distribution function~\cite{verma1972phonon,gofryk2014anisotropic}. 

\begin{figure}
    \centering
    \includegraphics[width=0.5\textwidth]{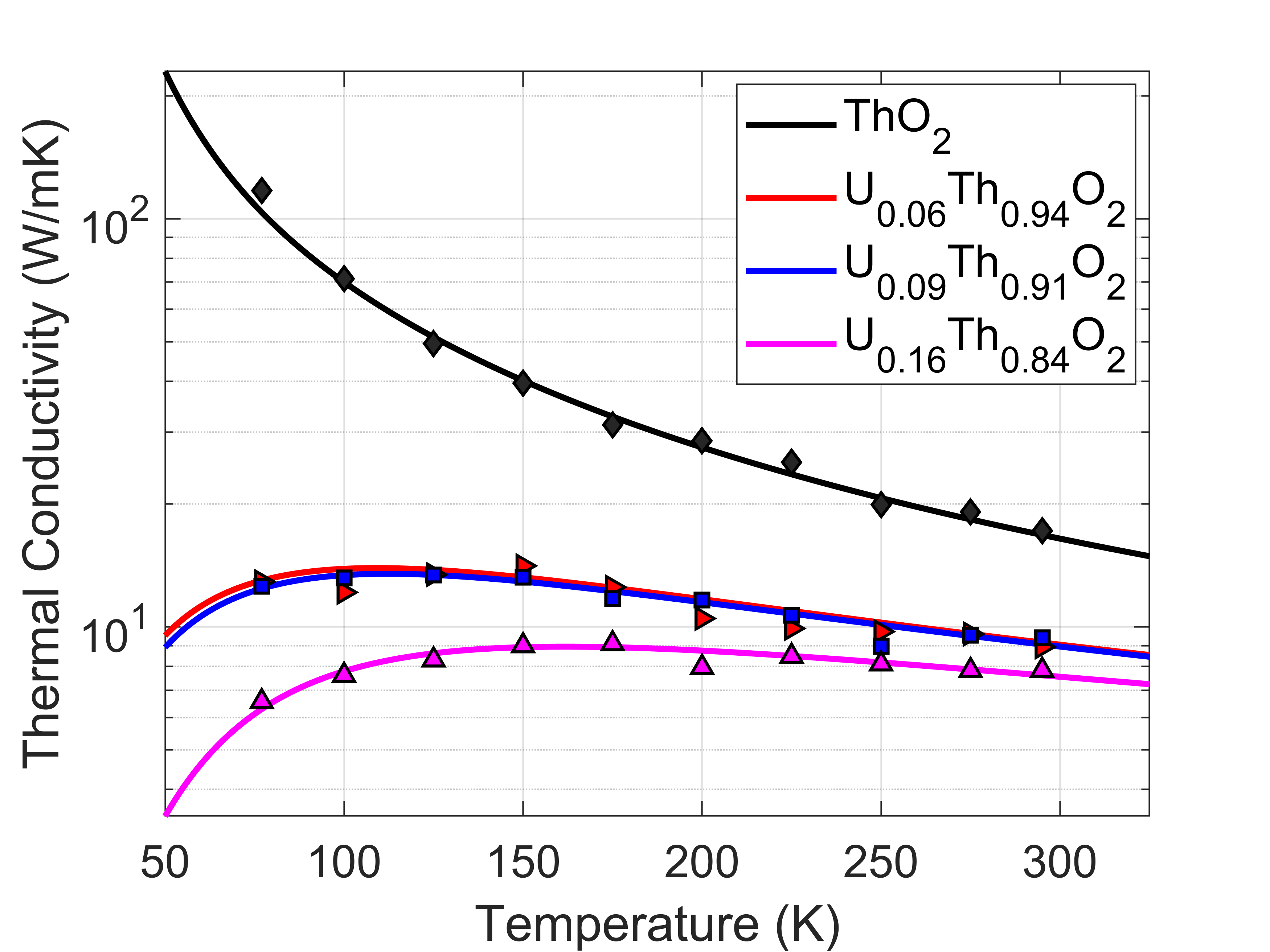}
    \caption{KCM fit of the temperature dependent thermal conductivity profiles of pristine and all U-doped ThO$_2$. Markers show the experimentally measured thermal conductivity values and solid lines the best fit.}
    \label{fig:uTho2_fit}
\end{figure}

The solid black line in~\cref{fig:fitting_options} (right) shows the fully optimized KCM taking into account both Rayleigh-type and resonant point defect scattering for the 16\% U-doped region, which clearly captures the temperature dependence of the thermal conductivity. Furthermore, comparing with Rayleigh-type scattering (dashed green line), resonant phonon scattering (dot-dashed red line) impacts thermal conductivity to the same magnitude in the temperature region above 250~K and more significantly below. In this optimization, $C_r$ and $\Gamma$ were used as local fitting parameters and $\omega_0$ as a global fitting parameter for all three doping levels simultaneously, while $B$ and $A_i$ are determined from the seed region and held constant. The best-fit KCM parameters for all samples are summarized in Table~\ref{tab:fitparameters}, with the optimized $k$--$T$ profiles for all three U-doped thoria regions presented in Fig.~\ref{fig:uTho2_fit}. Clearly, not only does the optimized KCM provide a qualitative match to the experimental results, best-fit values for $\Gamma$ and $C_r$ also scale monotonically with the U doping level, as would be expected if additional scattering is uranium concentration driven. However, we find that $C_r$, which captures the magnitude of resonant scattering effect scales non-linearly with uranium concentration, vary little between the 6\% and 9\% doping levels. Nevertheless, these features strongly suggest that resonant phonon scattering effects play a significant role in the U$_x$Th$_{1-x}$O$_2$ system.

\begin{table}
\centering
	\label{tab:fitparameters}
	\begin{tabular}{|l|c|c|c|c|}
		\hline
		 & ThO$_2$ & U$_{0.06}$Th$_{0.94}$O$_2$ & U$_{0.09}$Th$_{0.91}$O$_2$  & U$_{0.16}$Th$_{0.84}$O$_2$      \\ \hline
		B  {[}s/K{]}     & \multicolumn{4}{c}{$3.1\times10^{-18}$} \vline \\ \hline
		A$_i$  {[}s$^3${]}     & \multicolumn{4}{c}{$6.51\times10^{-44}$} \vline \\ \hline
		$\Gamma$         & $-$  & $0.0204$   &$0.0206$   &  $0.0214$  \\ \hline
		C$_r$  {[}s$^{-1}${]}       & $-$   &  $1.15\times10^{11}$   & $1.29\times10^{11}$  & $5.89\times10^{11}$  \\ \hline
		$\omega_0$  {[}THz{]}   & $-$  & \multicolumn{3}{c}{$3.83$} \vline \\ \hline
	\end{tabular}
	\caption{Optimized KCM fitting parameters for all U-doping levels measured. For doped regions, $\omega_0$ is globally optimized for all conductivity data simultaneously.}
\end{table}

While resonant phonon scattering related to the electronic structure of uranium has been previously studied, the measurements presented here provide the most direct evidence to date of the impact of this mechanism on phonon-mediated thermal transport. Doping uranium into the host thoria lattice introduces 5$f$ electrons not found in the ground state of thorium, the multipoles of which have been previously reported to interact strongly with phonons~\cite{Santini2009}. The presence of resonant phonon scattering in pure UO$_2$ has been confirmed by measuring the thermal conductivity of single crystal and polycrystalline UO$_2$ down to the liquid helium temperature~\cite{Moore1971,gofryk2014anisotropic}. Comparing to the $k$--$T$ profile measured on thoria with the same fluorite structure, these studies concluded that phonon scattering from the 5$f$ electronic degrees of freedom is responsible for the significantly lower thermal conductivity in UO$_2$. 



We can compare the extracted parameters of the phonon transport model in this study to the ones presented by Gofryk and coworkers on pure UO$_2$ for insight into the resonant scattering process~\cite{gofryk2014anisotropic}. In contrast to the model used here, Gofryk's work considered a two energy-level system to describe the spins of uranium ions, therefore with multiple resonances. The phonon resonant energy, converted from the resonant frequency $\omega_0$, is comparable: $\sim$2.5 meV here versus $\sim$3.5 meV for one of the UO$_2$ resonances reported previously. The resonant scattering intensity $C_r$, named the phonon-spin coupling constant in Gofryk’s work, is also at the same order of magnitude: up to $\sim11\times10^{11}$ s$^{-1}$ for both resonances combined, doubling the value measured on the 16\% U-doped thoria ($5.89\times10^{11}$ s$^{-1}$). The smaller-than-expected difference between the lightly-U-doped and pure UO$_2$ cases suggests that this resonant scattering process has a nonlinear dependence on doping level for high levels of U doping. Otherwise, the ratio observed here would suggest a saturation in $k$ reduction with $\sim$30-35\% U-doping. We caution that only qualitative comparisons should be made between the bulk single crystal thermal conductivity measured in this work and the micro-crystal measurements made by Gofryk and coworkers. 
In the latter case, boundary scatting effects known to significantly impact low temperature thermal conductivity must be explicitly accounted for~\cite{gofryk2014anisotropic} and radiative losses may also impact the as-measured values. In the case of the the millimeter-scale single crystals measured here, using thermal waves localized to linear distances in the tens of microns, boundary scattering should play no significant role~\cite{Dennett2021ThO2}.

The nonlinear dependence of resonant scattering on doping level can be extended to the low U doping region. With low U doping, differences between thermal transport in our 6\% and 9\% U-doped thoria are found negligible, but the difference between 6\% U-doped thoria and pure thoria is significant, indicating an extreme sensitivity to the presence of any U-doping-induced resonant scattering. To further explore the effects of very low concentrations of 5$f$ electrons, future studies should expand to samples with lower U doping levels. In related studies of other material systems, phonon resonant scattering has been observed at much lower impurity concentrations than studied here (e.g., 1$\times$10$^{19}$~cm$^{-3}$ in Mn-doped GaN~\cite{Begheri2020}, and Ni- and Cr-doped ZnSe~\cite{Lonchakov2004}, equivalently $\sim$0.1-0.5\%). These values provide a reasonable reference for targeted doping percentage ranges for future investigation.

\section{Conclusion}

In this work, the thermal conductivity of 6\%, 9\%, and 16\% U-doped thoria was experimentally measured in the temperature range 77--300~K. Measurements of a heterogeneous single crystal of U-doped ThO$_2$ were enabled by the isolation of micron-scale isoconcentration regions and local thermal conductivity measurements using SDTR. The measured thermal conductivity was compared to calculations from a Klemens-Callaway conductivity model to investigate the phonon scattering mechanisms in detail. Phonon resonant scattering induced by U doping is strongly suggested and its effects are found to be more impactful on thermal conductivity than Rayleigh-type point defect scattering. Comparing the fitted parameters of our model to those reported in a similar material system with different doping percentages and in pure UO$_2$, we hypothesize that the relationship between the resonant scattering intensity and the doping percentage is not linear, and it is likely that significant thermal conductivity reduction would appear with very low uranium doping percentages. 

\begin{acknowledgments}

This work was supported by the Center for Thermal Energy Transport under Irradiation (TETI), an Energy Frontier Research Center funded by the US Department of Energy, Office of Science, Office of Basic Energy Sciences. The isoconcentration region analysis work was supported by the Air Force Research Laboratory under award FA807518D0015.

\end{acknowledgments}

\section*{Data Availability Statement}

The data that support the findings of this study are available from the corresponding author upon reasonable request.

\nocite{*}
\bibliography{aipsamp}

\end{document}


\title[]{Supplementary Materials:\\ ~\vspace{-0.75pc} \\Thermal conductivity suppression in uranium-doped thorium dioxide due to phonon resonant scattering}

\author{Zilong Hua}
    \email{zilong.hua@inl.gov}
    \affiliation{Materials Science and Engineering Department, Idaho National Laboratory, Idaho Falls, ID 83415, USA}
\author{Saqeeb Adnan}
    \affiliation{Department of Mechanical and Aerospace Engineering, The Ohio State University, Columbus, OH 43210, USA}
\author{Amey R. Khanolkar}
    \affiliation{Materials Science and Engineering Department, Idaho National Laboratory, Idaho Falls, ID 83415, USA}
\author{Karl Rickert}
    \affiliation{KBR, Dayton, OH 45431, USA}
\author{David B. Turner}
    \affiliation{Azimuth Corporation, Fairborn, OH 45324, USA}
\author{J. Matthew Mann}
	\affiliation{Air Force Research Laboratory, Sensors Directorate, Wright-Patterson AFB, OH 45433, USA}
\author{David H. Hurley}
    \affiliation{Materials Science and Engineering Department, Idaho National Laboratory, Idaho Falls, ID 83415, USA}
\author{Marat Khafizov}
    \email{khafizov.1@osu.edu}
	\affiliation{Department of Mechanical and Aerospace Engineering, The Ohio State University, Columbus, OH 43210, USA}
\author{Cody A. Dennett}
	\affiliation{Materials Science and Engineering Department, Idaho National Laboratory, Idaho Falls, ID 83415, USA}
	\affiliation{Department of Nuclear Science and Engineering, Massachusetts Institute of Technology, Cambridge, MA 02139, USA}

\maketitle

\section*{Raman mapping of U-doped ThO$_2$ crystal}

Previous investigations of the Raman spectra of the U$_{1-x}$Th$_x$O$_2$ series establish that there is a single peak present for low U (high $x$) content, but additional peaks appear as the U content increases~\cite{Rickert2019,Rao2014,Bohler2014,Lee2018}. When a single peak is present, as is the case here, the peak corresponds to the $T_{2g}$ mode of the $Fm\bar{3}m$ space group~\cite{Keramidas1973}. With only a single peak being present, the peak intensity is of limited value as there is not a clear mechanism for normalization 
and differences in crystal thickness, surface features, and mounting can result in alterations to the measured intensity. To overcome this, previous reports have instead focused on the location or FWHM of the $T_{2g}$ peak. Of the two, the FWHM has proven to be more sensitive and less noisy and is provided in Figure 1(a) of the main manuscript. As can be observed, the FWHM of the ThO$_2$ is fully uniform and the seed/growth interface is discrete within the 1~{\textmu}m resolution of the instrumentation. The growth layer, however, is not fully homogenous. The FWHM closest to the seed (but not the interface itself) is the narrowest of the growth. This is suggestive of a high quality crystal with a low defect density. Since the incorporation of U would act as a defect in the pure ThO$_2$ structure, this also suggests that the U content is lowest at the seed/growth interface. Based solely upon the FWHM, the composition has two relatively uniform stripes that follow the interface for a short distance. After these a much larger degree of variation is present. That being said, there are certainly some areas that appear to have a local uniformity. The FWHM metric only evaluates the crystal quality, however. Any deductions made from these data could be purely the result of U incorporation or equally could be the consequences of some other kind of defect. 

\section*{XRF mapping of U-doped ThO$_2$ crystal}

Although the resolution of XRF is inferior to that of the {\textmu}-Raman spectroscopy (14.7~{\textmu}m vs. 1~{\textmu}m, respectively), it offers the direct interrogation as to whether the areas identified with the FWHM have a uniform composition or not. The same area that is interrogated with {\textmu}-Raman spectroscopy is subjected to XRF analysis as given in Figure 1(b) in the main manuscript. Three areas are identified that appear to be homogeneous from a qualitative inspection of both Raman and XRF maps. The quantitative compositions of these areas are listed in Figure 1(d) in the main manuscript. Some variation is present, but it is within the anticipated instrumental limits. For these three areas, having approximate U concentration of 16, 9, and 6 atomic percent, the U and Th content can be treated as uniform.
\bibliography{ref}